\documentclass[12pt]{article}
\usepackage{epsfig, amssymb}
\usepackage{graphicx,epsfig}
\usepackage{epstopdf}
\begin{document}
\begin{titlepage}
\thispagestyle{empty}
\begin{flushright}
\end{flushright}

\bigskip

\begin{center}
\noindent{\Large \textbf
{Boundary conditions for conformally coupled scalar in AdS$_4$}}\\ 
\vspace{2cm} \noindent{
Jae-Hyuk Oh\footnote{e-mail:jack.jaehyuk.oh@gmail.com}}

\vspace{1cm}
  {\it
Department of Physics, Hanyang University, Seoul 133-791, Korea\\
 }
\end{center}

\vspace{0.3cm}
\begin{abstract}
We consider conformally coupled scalar with $\phi^4$ coupling in AdS$_4$ and study  its various boundary conditions on AdS boundary. We have obtained perturbative solutions of equation of motion of the conformally coupled scalar with power expansion order by order in $\phi^4$ coupling $\lambda$ up to $\lambda^2$ order. In its dual CFT, we get 2,4 and 6 point functions  by using this solution with Dirichlet and Neumann boundary conditions via AdS/CFT dictionary. We also consider  
marginal deformation on AdS boundary and get its on-shell and boundary effective actions.
\end{abstract}
\end{titlepage}

\tableofcontents
\section{Introduction}
Alternative quantization is studied by Breitenlohner and Freedman\cite{BF} in gauged supergravity context and Klevanov and Witten et. al. \cite{Klebanov:1999tb,Witten:2001ua,Sever:2002fk} discussed this in AdS/CFT which gives interesting conformal field theories as duals of
the gravity theories defined in AdS space by imposing various possible boundary conditions on AdS boundary. 

For scalar field theories in AdS space\cite{de Haro:2006nv,deHaro:2006wy}, the dual conformal field theory is known to be unitary when its mass is in the region of $-\frac{d^2}{4}\leq m^2\leq -\frac{d^2}{4}+1$, where $d$ is dimensionality of AdS boundary and $m$ is mass of the scalar field. In this mass window, there are two possible quantization schemes in the dual CFT, which are $\Delta_+$ and $\Delta_-$ theories respectively, where $\Delta_{\pm}=\frac{d}{2}\pm\sqrt{\frac{d^2}{4}+m^2}$, the conformal dimensions of the boundary operator in each dual CFT. Since the unitarity bound of the scalar operators in CFT is $\Delta\geq \frac{d}{2}-1$, two point correlation functions in both CFTs are positive definite
\footnote{There are many discussions about alternative quantization schemes in Dirac field\cite{Henneaux:1998ch}, Rarita-Schwinger field\cite{Rashkov:1999ji}, $U(1)$ gauge field\cite{deHaro:2007eg} and $SU(2)$ Yang-Mills field\cite{Jatkar:2012mm}.}.

The bulk scalar field shows near AdS boundary expansion as $\phi(r,x_i)=\phi^{(0)}(x_i)r^{d-\Delta_+}+\phi^{(1)}(x_i)r^{\Delta_+}$ as the AdS radial coordinate $r\rightarrow0$, where $\phi^{(0)}(x_i)$ becomes a source term in the dual CFT whereas $\phi^{(1)}(x_i)$ is certain boundary composite operator according to standard AdS/CFT dictionary. This is called $\Delta_+$ theory and achieved by imposing Dirichlet boundary condition $\delta\phi^{(0)}=0$. This gives standard quantization scheme to the dual CFT. Even in the case that the scalar field mass in not in the mass range, $\Delta_+$ satisfies unitarity bound and so Dirichlet boundary condition is always possible boundary condition. 

Alternative quantization can be achieved by imposing Neumann boundary condition as $\delta \phi^{(1)}=0$. In this quantization scheme, the role of the source and composite operator in the dual CFT is switched, so $\phi^{(0)}(x_i)$ becomes the boundary composite operator and $\phi^{(1)}(x_i)$ is the source term.
The corresponding CFT is called $\Delta_-$ theory.

A way of imposing boundary conditions is to add boundary term on AdS boundary and variational principle provides the boundary condition upto bulk equation of motion. Therefore, once we define on-shell action  as $I_{os}=S_{bulk}+S_{bdy}$ and then the boundary condition is $\delta I_{os}=0$, where $S_{bulk}$ is boundary contribution of the bulk scalar field action upto the equation of motion and $S_{bdy}$ is boundary action. Dirichlet boundary condition is achieved by $S_{bdy}=0$. Without adding any boundary term, $\delta I_{os}=\int \phi^{(1)}\delta \phi^{(0)}=0$, so we can request $\delta \phi^{(0)}=0$. Neumann boundary condition is obtained by adding $S_{bdy}=-\int \phi^{(0)}\phi^{(1)}$ to the bulk action. Then $\delta I_{os}=-\int \phi^{(0)}\delta\phi^{(1)}$, so the boundary condition becomes $\delta \phi^{(1)}=0$.

In this paper, we study conformally coupled scalar field $\phi$ in AdS$_4$ with $\phi^4$ interaction. Mass of the conformally coupled scalar is
$m^2=-\frac{d^2-1}{4}$. Therefore in AdS$_4$, $m^2=-2$ so $\Delta_+=2$ and $\Delta_-=1$ which satisfy unitarity bound. Definitely two possible quantization schemes can be applied to the conformally coupled scalar and there are various boundary CFTs obtained by imposing various kinds of boundary conditions. One interesting property of this conformally coupled scalar is that a suitable field redefinition transforms this theory into a massless scalar field theory in half of flat space, $\mathbb R_{+}$. In our study, we mostly work in this $\mathbb R_+$ frame.

Conformally coupled scalar in AdS and their boundary CFTs are studied in several literatures. In \cite{Papadimitriou:2007sj}, the author have considered conformally coupled scalar in AdS space and their boundary effective action with two derivative kinetic term by using derivative expansion(i.e. by ignoring higher derivative terms). In \cite{de Haro:2006nv,deHaro:2006wy,Papadimitriou:2007sj}, the authors have considered conformally coupled scalar and its instanton solution. By looking at their solutions, they suggested that the dual CFT theory shows $\phi^6$ interaction with standard(two derivative) kinetic term.

In these studies, however, the authors do not consider small coupling regime of the dual CFTs. 
Derivative expansion provides basically the low momentum regime of the theory  ignoring higher derivative terms. 
Therefore, in \cite{Papadimitriou:2007sj}, they obtained low energy effective action of the dual field theory.
Moreover, instanton solution cannot probe perturbative regime of theories either.

In this note, we concentrate on the perturbative regime in the dual CFTs.
We have obtained the boundary effective action in dual field theories with its exact 2-point propagator and 3,4 point interaction vertices in the perturbative regime with power expansion order by order in the small coupling $\lambda$, which is $\phi^4$ interaction coupling in the dual gravity theory. 
The couplings of multi point functions in the dual CFTs are written in terms of a certain power of the coupling $\lambda$.
 In fact, it turns out that the strength of the $2n+2$ point function  in the boundary effective theory is proportional to $\lambda^n$.

We address the main results in detail.
In section \ref{Conformally coupled scalar in AdS$_4$ and its perturbative solutions}, we solve conformally coupled scalar field equation of motion with power expansion order by order in $\lambda$. 
We obtain the solutions of the equation of motion upto $\lambda^2$ order. 
In section \ref{Boundary conditions and effective actions}, by utilizing the bulk scalar field solutions, we compute on-shell and boundary effective actions by imposing several different boundary conditions. For Dirichlet boundary condition, we obtained 2,4  and 6-point functions in dual CFTs. 
2-point function(propagator) of the dual composite operators is given by
\begin{equation}
\langle O_pO_q \rangle=-|p|\delta^3(p+q),
\end{equation}
which is proportional to the absolute value of 3-momentum $p$ along the boundary directions.
4-point function is given by
\begin{equation}
\langle O_pO_{q}O_sO_{u}\rangle=-\frac{3!\lambda}{(2\pi)^3}\frac{\delta^3(u+q+s+p)}{|u|+|q|+|s|+|p|},
\end{equation} 
where $u$, $q$, $s$ and $p$ are 3-momenta along boundary directions too and $O$ indicates the dual CFT operator. This is exotic and non local boundary momentum dependent correlation function. 
6-point function is too complex to address here, which is also external momenta dependent. 

For Neumann boundary condition, the 4-point function of boundary composite operator becomes
\begin{equation}
\langle \phi^{(0)}_{p}\phi^{(0)}_{q}\phi^{(0)}_{s}\phi^{(0)}_{u}\rangle=\frac{\langle O_pO_{q}O_sO_{u}\rangle}{|u||q||s||p|},
\end{equation}
where $\phi^{(0)}_p$ is boundary value of the bulk scalar field $\phi$ and it becomes the dual field theory operator in Neumann boundary condition.

We also consider another type of boundary condition, called marginal deformation\cite{de Haro:2006nv} and obtained its boundary effective action too.

\section{Conformally coupled scalar in AdS$_4$ and its perturbative solutions}
\label{Conformally coupled scalar in AdS$_4$ and its perturbative solutions}
In this section,  we solve the equation of motion of the conformally coupled scalar with $\phi^4$ interaction term with power expansion order by order in its coupling $\lambda$.   
We start with the action
\begin{equation}
\label{scalar-in-ads-action}
S=\int drd^3x\sqrt{g}\mathcal L(\phi) + \int d^3x \mathcal L_{c.t.}(\phi)+S_{bdy},
\end{equation}
where $\mathcal L$ is an action of conformally coupled scalar in AdS$_4$, which is given by
\begin{equation}
\mathcal L= \frac{1}{2}g^{\mu\nu}\partial_\mu \phi \partial_\nu \phi-\phi^2+\frac{\lambda}{4}\phi^4,
\end{equation}
$\mathcal L_{c.t.}$ is counter term Lagrangian and $S_{bdy}=\int d^3x \mathcal L_{b}$ is boundary action. (Euclidean) AdS$_4$ metric is given by
\begin{equation}
\label{AdS-metric}
ds^2=g_{\mu\nu}dx^\mu dx^\nu=\frac{dr^2+\sum_{i=1}^3 dx^i dx^i}{r^2},
\end{equation}
where $r=x^0$ is the radial coordinate of AdS space($r$=0 is AdS boundary and $r=\infty$ is Poincaré horizon), $x^i$ is boundary directional coordinate and the spacetime indices $i$, $j$... run from 1 to 3. The action of conformally coupled scalar enjoys an interesting property as follows. Once we define a new field $f(r,x)=\frac{\phi(r,x)}{r}$, then by using the explicit form of the background metric(\ref{AdS-metric}), the action(\ref{scalar-in-ads-action}) is transformed into
\begin{equation}
\label{f-action}
S=\int drd^3x \left(\frac{1}{2}\partial_r f \partial_r f + \frac{1}{2} \delta^{ij}\partial_i f \partial_j f +\frac{\lambda}{4}f^4\right)+\int d^3x\left(\mathcal L_{c.t.}(\phi)+\frac{f^2}{2r}\right)+\int d^3x \mathcal L_{b}(\phi),
\end{equation}
where the $\frac{f^2}{2r}$ term may divergent as it approach AdS boundary. Once we take $\mathcal L_{c.t.}=-\sqrt{\gamma}\frac{\phi^2}{2}$, then this term is canceled with $\mathcal L_{c.t.}$, where $\gamma$ is determinant of an induced metric $\gamma_{ij}$ from $g_{\mu\nu}$ as
\begin{equation}
\gamma_{ij}=\frac{\partial x^\mu}{\partial x^i}\frac{\partial x^\nu}{\partial x^j}g_{\mu\nu}.
\end{equation} 
Equation of motion is obtained by variation of the action(\ref{f-action}), which is given by
\begin{equation}
0=\partial^2_r f +\delta^{ij}\partial_i \partial_j f -\lambda f^3.
\end{equation}
By using Fourier transform,
\begin{equation}
\label{Fourier-transform}
f(x)=\frac{1}{(2\pi)^{\frac{3}{2}}}\int e^{-ip_i x_i}f_p(r) d^3p
\end{equation}
one can get this equation in momentum space:
\begin{equation}
\label{equation of motion in momentum space}
0=\partial^2_r f_p(r)-p^2f_p(r)-\frac{\lambda}{(2\pi)^3}\int d^3[q,s,t]\delta^3(q+s+t-p)f_q(r)f_s(r)f_t(r),
\end{equation}
where $d^3[q,s,...,t]\equiv d^3qd^3s...d^3t$
\footnote{We will use $p,q,s,t,u,v,w$ to indicate 3-momenta along boundary directions.}. 

We solve the equation perturbatively order by order in $\lambda$ upto $\lambda^2$ order. Namely, we try the following ansatz:
\begin{equation}
f_p(r)=\bar f_p(r)+\lambda \tilde f_p(r)+\lambda^2 \hat f_p(r)+O(\lambda^3)
\end{equation}
In the zeroth order in $\lambda$, the equation becomes
\begin{equation}
0=\partial^2_r \bar f_p(r)-p^2\bar f_p(r),
\end{equation}
and its solution is given by
\begin{equation}
\bar f_p(r)=\bar f_{0,p}\cosh(|p|r)+\frac{\bar f_{1,p}}{|p|}\sinh(|p|r),
\end{equation}
where $\bar f_{0,p}$ and $\bar f_{1,p}$ are boundary momenta, $p_i$ dependent functions and $|p|=\sqrt{p^2_1+p^2_2+p^2_3}$, which are absolute value of momentum along boundary direction. This solution should be regular everywhere, and for this we request that
\begin{equation}
\bar f_{0,p}+\frac{\bar f_{1,p}}{|p|}=0,
\end{equation}
Thus, the regular solution is given by
\begin{equation}
\bar f_p(r)=\bar f_{0,p}e^{-|p|r}.
\end{equation}

In the first order in $\lambda$, the equation is given by
\begin{equation}
0=(\partial^2_r-p^2)\tilde f_p(r)-\frac{1}{(2\pi)^3}\int d^3[q,s,t]\delta^3(q+s+t-p)\bar f_{0,q}\bar f_{0,s}\bar f_{0,t}e^{-(|q|+|s|+|t|)r},
\end{equation}
where the last term is a source term from the zeroth order solution. The first order solution is given by
\begin{equation}
\tilde f_p(r)=\tilde f_{0,p}e^{-|p|r}+\frac{1}{(2\pi)^3}\int d^3[q,s]f_{0,p-q-s}f_{0,q}f_{0,s}\frac{e^{-(|p-q-s|+|q|+|s|)r}}{(|p-q-s|+|q|+|s|)^2-p^2},
\end{equation}
where the first term is homogeneous solution and the last term is inhomogeneous one. 

Finally, we will obtain the second order solution in $\lambda$. The equation of motion is given by
\begin{eqnarray}
0=(\partial^2_r-p^2)\hat f_p(r)-\frac{1}{(2\pi)^6}\int d^3[t,q,s,v,u]\delta^3(t+q+s-p)\\ \nonumber \times f_{0,t}f_{0,q}f_{0,u}f_{0,v}f_{0,s-u-v}\frac{e^{-(|t|+|q|+|u|+|v|+|s-u-v|)r}}{(|s-u-v|+|u|+|v|)^2-s^2},
\end{eqnarray}
and its solution becomes
\begin{eqnarray}
\hat f_p(r)&=& \hat f_{0,p}e^{-|p|r}+ \frac{3}{(2\pi)^6}\int d^3[t,q,s,u,v,w]f_{0,t}f_{0,q}f_{0,v}f_{0,u}f_{0,w} \\ \nonumber
&\times&\frac{\delta^3(t+q+s-p)\delta^3(w+u+v-s)}{[(|w|+|v|+|u|)^2-s^2][(|w|+|t|+|q|+|v|+|u|)^2-p^2]}e^{-(|w|+|t|+|q|+|v|+|u|)r},
\end{eqnarray}
where again the first term in the solution is homogeneous part and the last term is inhomogeneous one.

The homogeneous solutions in the first and the second order in $\lambda$ can be absorbed in the zeroth order solution. Therefore, we set $\tilde f_{0,p}=\hat f_{0,p}=0$. Moreover, we define a few complex expressions as
\begin{eqnarray}
\alpha_p(u,q,s)&\equiv&\frac{\delta^3(u+q+s-p)}{(|u|+|q|+|s|)^2-p^2}, \\
\beta_p(t,q;v,s,u)&\equiv&\frac{\delta^3(q+u+s+t+v-p)}{[(|v|+|s|+|u|)^2-(v+s+u)^2]}\\ \nonumber
&\times& \frac{1}{[(|v|+|t|+|q|+|s|+|u|)^2-p^2]}
\end{eqnarray}
then, the form of the solution is much more simplified.

\paragraph{Near boundary expansion}
Near conformal boundary $r=0$, our solution is expanded as
\begin{eqnarray}
f_p(r)&=&f_{0,p}+\frac{\lambda}{(2\pi)^3}\int d^3[q,s,u]f_{0,q}f_{0,s}f_{0,u}\alpha_p(q,s,u)\\ \nonumber
&+&\frac{3\lambda^2}{(2\pi)^6}\int d^3[t,q,s,u,v]f_{0,q}f_{0,s}f_{0,u}f_{0,t}f_{0,v}\beta_p(t,q;v,s,u) \\ \nonumber
&+&r\left[ -|p|f_{0,p}- \frac{\lambda}{(2\pi)^3}\int d^3[q,s,u]f_{0,q}f_{0,s}f_{0,u}(|q|+|s|+|u|)\alpha_p(q,s,u)\right. \\ \nonumber
&-&\left.\frac{3\lambda^2}{(2\pi)^6}\int d^3[t,q,s,u,v]f_{0,q}f_{0,s}f_{0,u}f_{0,t}f_{0,v}(|v|+|t|+|q|+|s|+|u|)\beta_p(t,q;v,s,u)\right]+O(r^2).
\end{eqnarray}
The first two lines are boundary value of the field $f_p(r)$ while the third and fourth lines are the boundary value of $\partial_r f_p(r)$. We define this boundary value of $f_p(r)$ to be $f^{(0)}_p$ and the boundary value of $\partial_r f_p(r)\equiv f^{(1)}_p$. One can rewrite $f^{(1)}_p$ in terms of
the boundary value $f^{(0)}_p$, which is given by
\begin{eqnarray}
f^{(1)}_p&=&-|p|f^{(0)}_p-\frac{\lambda}{(2\pi)^3}\int d^3[q,s,u]f^{(0)}_u f^{(0)}_q f^{(0)}_s \alpha_p(u,q,s)(|u|+|q|+|s|-|p|) \\ \nonumber
&+&\frac{3\lambda^2}{(2\pi)^6}\int d^3[q,u,t,v,w]f^{(0)}_uf^{(0)}_qf^{(0)}_tf^{(0)}_vf^{(0)}_w[\alpha^{(2)}_p(u,q;t,v,w)(|u|+|q|+|t+v+w|-|p|)\\ \nonumber
&-&\beta_p(t,q;v,w,u)(|v|+|t|+|q|+|w|+|u|-|p|)],
\end{eqnarray}
where
\begin{eqnarray}
\alpha^{(2)}_p(u,q;t,v,w)&\equiv&\int d^3s \alpha_p(u,q,s)\alpha_s(t,v,w)\\ \nonumber
&=&\frac{\delta^3(t+v+w+q+u-p)}{[(|t|+|v|+|w|)^2-(t+v+w)^2][(|t+v+w|+|q|+|u|)^2-p^2]}
\end{eqnarray}

\section{Boundary conditions and effective actions}
\label{Boundary conditions and effective actions}
In this section, we discuss various boundary conditions and on-shell and boundary effective actions followed by those boundary conditions. One can evaluate on-shell action by using equation of motion from (\ref{f-action}). This is given by
\begin{equation}
I_{os}=S_{bulk}+S_{bdy}=\int_{r=0} d^3x\frac{1}{2}f(x,r)\partial_r f(x,r)-\int drd^3x\frac{\lambda}{4}f^4(x,r)+\int d^3x \mathcal L_b(f).
\end{equation}
By using Fourier transform defined in (\ref{Fourier-transform}),
one can write this in momentum space as
\begin{eqnarray}
\label{bulk action in momentum space}
S_{os}=\frac{1}{2}\int_{r=0} d^3 pf_p(r)\partial_r f_p(r)-\frac{\lambda}{4(2\pi)^3}\int_{r=0}d^3 [p,q,s,t] dr f_p(r)f_q(r)f_s(r)f_t(r)\delta^3(p+q+s+t)\\ \nonumber
+\int d^3p \mathcal L_b(f_p).
\end{eqnarray}

We define the boundary value of the bulk canonical momentum, $\partial_r f(r)$ as
\begin{equation}
\label{canonical momentum}
\Pi_{-p}=\frac{\delta S_{bulk}}{\delta f^{(0)}_p}.
\end{equation}
With this canonical momentum, the on-shell action is now functional of the boundary value of $f_p(r)$, $f^{(0)}_p$ and its canonical momentum $\Pi_{-p}$. In AdS/CFT context, the bulk on-shell action becomes generating functional of the dual CFT as
\begin{equation}
Z[J]=e^{-W[J(f^{(0)}_p,\Pi_{-p})]}=\int D[f^{(0)}_p,\Pi_{-p}]\exp\left[-S_{bulk}(f^{(0)}_p)-S_{bdy}(f^{(0)}_p,\Pi_{-p})  \right],
\end{equation}
where $J$ is source which couples to certain boundary composite operator and $W$ is the generating functional with the source term $J$. This generating functional is identified with the on-shell action as $W[J]=I_{os}[f^{(0)}_p,\Pi_{-p}]$. In standard quantization($\Delta_+$ theory), $J=f^{(0)}_p$ and the boundary composite operator becomes $\Pi_{-p}$. In general, however,  the source $J$ is generic function of $f^{(0)}_p$ and $\Pi_{-p}$.

The boundary condition is obtained by looking at saddle point of the on-shell action as
\begin{equation}
\frac{\delta I_{os}[f^{(0)}_p,\Pi_{-p}]}{\delta f^{(0)}_p}=0, {\rm\ \ and\ \ }
\frac{\delta I_{os}[f^{(0)}_p,\Pi_{-p}]}{\delta \Pi_{-p}}=0,
\end{equation}
which gives the relation between $f^{(0)}_p$ and $\Pi_{-p}$ and with this, one can rewrite the on-shell action in terms of $f^{(0)}_p$ only, which also gives the correct boundary condition in its saddle point.

The boundary effective action is given by Legendre transform of the generating functional, which if given by
\begin{equation}
\label{legendre transform}
\Gamma[\sigma]=-\int J\sigma + W[J],
\end{equation}
where $\sigma$ is vacuum expectation value of the boundary composite operator. 
Followed by (\ref{legendre transform}), $\sigma$ and $J$ satisfy the following relations:
\begin{equation}
\sigma=\frac{\delta W[J]}{\delta J}{\rm \ \  and \ \ }J=-\frac{\delta\Gamma[\sigma]}{\delta\sigma}.
\end{equation}
Suppose that $W$ and $\Gamma$ are the generating functional and boundary effective action without any boundary deformation. Let us deform this with boundary term $S_{bdy}$, and assume that $\sigma$ does not change by the deformation. The deformed boundary effective action may have a form of
\begin{equation}
\Gamma_{d}[\sigma]=\Gamma[\sigma]+\int g(\sigma),
\end{equation}
where $\Gamma_d$ is deformed boundary effective action and $g$ is a function of $\sigma$. Deformed source $J_d$ will be given by
\begin{equation}
J_d\equiv\frac{\delta \Gamma_d[\sigma]}{\delta \sigma}=J-\frac{dg(\sigma)}{d\sigma},
\end{equation}
Therefore, deformed generating functional $W_d[J_d]=\Gamma_d[\sigma]+\int J_d\sigma$ is given by
\begin{equation}
W_d[J_d]=W[J]+\int (g(\sigma)-\sigma g^\prime(\sigma)).
\end{equation}
In sum, the boundary deformation term is given by
\begin{equation}
S_{bdy}=\left.\int (g(\sigma)-\sigma g^\prime(\sigma))\right|_{\sigma=\frac{\delta W[J]}{\delta J}}.
\end{equation}

\paragraph{Dirichlet boundary condition}
Dirichlet boundary condition is achieved without adding any deformation term. Then, the boundary condition is obtained by finding saddle point of the bulk action as
\begin{equation}
\delta I_{os}=\delta S_{bulk}= \int \Pi_{-p}\delta f^{(0)}_p,
\end{equation} 
Then the boundary condition is $\delta f^{(0)}_p=0$.
By plugging the on-shell solution into (\ref{bulk action in momentum space}), one can evaluate this on-shell action in terms of the boundary value of the field $f(r)$ as

\begin{eqnarray}
I^D_{os}=W^D[f^{(0)}_p]=\frac{1}{2}\int d^3pd^3 qf^{(0)}_pf^{(0)}_{q}\langle O_pO_{q}\rangle+\frac{1}{4!}\int d^3 [p,q,s,u] f^{(0)}_pf^{(0)}_qf^{(0)}_sf^{(0)}_u\langle O_pO_{q}O_sO_{u}\rangle \\ \nonumber
+\frac{1}{6!}\int d^3 [p,q,t,v,w,u]
f^{(0)}_pf^{(0)}_qf^{(0)}_tf^{(0)}_vf^{(0)}_wf^{(0)}_u\langle O_pO_{q}O_tO_{v}O_wO_{u}\rangle
\end{eqnarray}
where
\begin{equation}
\langle O_pO_{q}\rangle=-|p|\delta^3(p+q),
\end{equation}
\begin{equation}
\langle O_pO_{q}O_sO_{u}\rangle=-\frac{3!\lambda}{(2\pi)^3}\frac{\delta^3(u+q+s+p)}{|u|+|q|+|s|+|p|},
\end{equation}
and
\begin{eqnarray}
\langle O_pO_{q}O_wO_{u}O_tO_{v}\rangle=\frac{9\cdot 5!\lambda^2}{(2\pi)^6}\left(\frac{\delta^3(t+v+w+q+u+p)}{[(|t|+|v|+|w|)^2-(t+v+w)^2][|t+v+w|+|u|+|q|+|p|]}\right.\\ \nonumber
-\left.\frac{\delta^3(t+v+w+q+u+p)}{[(|u|+|p|+|q|)^2-(t+v+w)^2][|t|+|v|+|w|+|u|+|q|+|p|]}\right. \\ \nonumber
-\left.\frac{2/3\ \delta^3(t+v+w+q+u+p)}{[|t|+|v|+|w|+|u|+|q|+|p|][|t|+|v|+|w|+|t+v+w|][|t+v+w|+|u|+|q|+|p|]}\right).
\end{eqnarray}
This provides boundary momentum dependent 2,4, and 6-point functions in dual CFTs.

From the definition(\ref{canonical momentum}), we get canonical momentum of $f^{(0)}$:
\begin{eqnarray}
 \Pi_{-p}=\int d^3sd^3 q\delta^3(s-p)f^{(0)}_{q}\langle O_sO_{q}\rangle+\frac{1}{3!}\int d^3 [t,q,s,u] \delta^3(t-p)f^{(0)}_qf^{(0)}_sf^{(0)}_u\langle O_tO_{q}O_sO_{u}\rangle \\ \nonumber
+\frac{1}{5!}\int d^3 [p,q,t,v,w,u]
f^{(0)}_pf^{(0)}_qf^{(0)}_tf^{(0)}_vf^{(0)}_wf^{(0)}_u\langle O_pO_{q}O_tO_{v}O_wO_{u}\rangle
\end{eqnarray}
By Legendre transform (\ref{legendre transform}),
Boundary effective action is given by
\begin{eqnarray}
\Gamma^D(\Pi)=-\frac{1}{2}\int d^3[p,q]\frac{\langle O_pO_{q}\rangle}{|p||q|}\Pi_{-p}\Pi_{-q}+\frac{1}{4!}\int d^3[p,q,s,t]\frac{\langle O_pO_{q}O_sO_{t}\rangle}{|p||q||s||t|}\Pi_{-p}\Pi_{-q}\Pi_{-s}\Pi_{-t}\\ \nonumber
+\frac{1}{3!}\int d^3[p,q,s,t,u,v]\Pi_{-p}\Pi_{-q}\Pi_{-s}\Pi_{-t}\Pi_{-u}\Pi_{-v}\left(\frac{1}{2}\int d^3w \frac{\langle O_pO_{q}O_sO_{w}\rangle}{|p||q||s||w|}|w|\frac{\langle O_wO_{t}O_uO_{v}\rangle}{|w||t||u||v|}\right. \\ \nonumber
+\left.\frac{1}{20}\frac{\langle O_pO_{q}O_sO_{t}O_uO_{v}\rangle}{|p||q||s||t||u||v|}\right),
\end{eqnarray}

\paragraph{Neumann boundary condition} 
Neumann boundary condition is achieved by adding boundary deformation to the bulk action as
\begin{equation}
S^N_{bdy}=-\int d^3p f^{(0)}_p \Pi_{-p},
\end{equation}
then the boundary condition is achieved by variation of the on-shell action
\begin{equation}
\delta I_{os}=\delta S_{bulk}+\delta S_{bdy}=-\int d^3 p f^{(0)}_p\delta \Pi_{-p}.
\end{equation}
Therefore, one can request $\delta\Pi_{-p}=0$, which is Neumann boundary condition. Since adding such boundary deformation provides effective Legendre transform from bulk action, then the on-shell action has the same form with the boundary effective action in Dirichlet boundary condition case. Moreover, its boundary effective action will be the form of the on-shell action in Dirichlet boundary condition case too. In sum,
\begin{equation}
I^N_{os}=W^N[\Pi_{-p}]=\Gamma^D[\Pi_{-p}], {\rm\ \ and\ \ }I^D_{os}=W^D[f^{(0)}_{p}]=\Gamma^N[f^{(0)}_{-p}].
\end{equation}

\paragraph{Marginal deformation} Marginal deformation is achieved by adding the following boundary deformation term:
\begin{equation}
S^M_{bdy}=-\frac{\alpha}{3}\int \frac{d^3[p,q,t]}{(2\pi)^{3/2}}f^{(0)}_pf^{(0)}_qf^{(0)}_t\delta^3(p+q+t),
\end{equation}
where $\alpha$ is a free real parameter.
Then, followed by this, the boundary condition
\footnote{Even in this case, Dirichlet boundary condition is possible to be imposed. Dirichlet boundary condition is always possible boundary condition.}
is given by
\begin{equation}
0=\frac{\delta S}{\delta f^{(0)}_p}=\Pi_{-p}-\frac{\alpha}{3}\int \frac{d^3[q,t]}{(2\pi)^{3/2}}
f^{(0)}_qf^{(0)}_t\delta^3(p+q+t)
\end{equation}

On-shell action in marginal deformation case is 
\begin{equation}
I^M_{os}=I^D_{os}[f^{(0)}_p]+S^M_{bdy}[f^{(0)}_p]
\end{equation}
By using the procedure introduced in the beginning of this section to derive boundary effective action from on-shell action and the source $J$, we get those as
\begin{eqnarray}
 \Gamma^M(f^{(0)}_p)=-\frac{1}{2}\int d^3pd^3 qf^{(0)}_pf^{(0)}_{q}\langle O_pO_{q}\rangle
 +\frac{1}{3}\int d^3[q,s,t]f^{(0)}_qf^{(0)}_sf^{(0)}_t\langle O_qO_sO_t \rangle \\ \nonumber
 -\frac{1}{3\cdot4!}\int d^3 [p,q,s,u] f^{(0)}_pf^{(0)}_qf^{(0)}_sf^{(0)}_u\langle O_pO_{q}O_sO_{u}\rangle \\ \nonumber
-\frac{1}{5\cdot6!}\int d^3 [p,q,t,v,w,u]
f^{(0)}_pf^{(0)}_qf^{(0)}_tf^{(0)}_vf^{(0)}_wf^{(0)}_u\langle O_pO_{q}O_tO_{v}O_wO_{u}\rangle,
\end{eqnarray}
\begin{eqnarray}
J^M_{-p}=\int d^3 qf^{(0)}_{q}\langle O_pO_{q}\rangle
 -\int d^3[s,t]f^{(0)}_sf^{(0)}_t\langle O_pO_sO_t \rangle \\ \nonumber
 +\frac{1}{3\cdot3!}\int d^3 [q,s,u] f^{(0)}_qf^{(0)}_sf^{(0)}_u\langle O_pO_{q}O_sO_{u}\rangle \\ \nonumber
+\frac{1}{5\cdot5!}\int d^3 [q,t,v,w,u]
f^{(0)}_qf^{(0)}_tf^{(0)}_vf^{(0)}_wf^{(0)}_u\langle O_pO_{q}O_tO_{v}O_wO_{u}\rangle,
\end{eqnarray}

where
\begin{equation}
\langle O_qO_sO_t \rangle=\frac{\alpha}{2}\frac{1}{(2\pi)^{3/2}}\delta^3({q+s+t}).
\end{equation}

\section*{Acknowledgement}
J.H.O would like his $\mathcal{W.J.}$ This work is supported by the research fund of Hanyang University(HY-2013) only.

\end{document}